\documentstyle[twoside,fleqn,espcrc2]{article}

\newcommand{\AmS}{{\protect\the\textfont2
  A\kern-.1667em\lower.5ex\hbox{M}\kern-.125emS}}

\newcommand{\be}{\begin{equation}}                                              
\newcommand{\ee}{\end{equation}}                                                
\newcommand{\half}{\frac{1}{2}}

\newcommand{\LCB}{\raisebox{-0.3ex}{\mbox{\LARGE$\left\{\right.$}}}
\newcommand{\RCB}{\raisebox{-0.3ex}{\mbox{\LARGE$\left.\right\}$}}}
\newcommand{\LSB}{\raisebox{-0.3ex}{\mbox{\LARGE$\left[\right.$}}}
\newcommand{\RSB}{\raisebox{-0.3ex}{\mbox{\LARGE$\left.\right]$}}}

\title{Supersymmetric gauge theories on the lattice}

\author{I. Montvay
\address{Deutsches Elektronen Synchrotron DESY, \\
         Notkestr. 85, D-22603 Hamburg, Germany}}
       
\begin{document}

\begin{abstract}
 The perspectives of numerical simulations in supersymmetric quantum
 field theories with vector-like gauge symmetries are discussed.
 A numerical simulation algorithm for SU(2) gauge theory with
 gluinos is studied and the first results on the glueball-gluinoball
 spectrum are presented. 
\end{abstract}

\maketitle

\section{INTRODUCTION}

 Supersymmetric gauge theories may have fascinating non-perturbative
 properties.
 Previous attempts to understand them theoretically are mainly based on
 the instanton calculus (for a review and references see \cite{AKMRV}).
 Recently, the $N=2$ extended gauge theories received particular
 interest because Seiberg and Witten were able to obtain exact results
 on their low energy effective action \cite{SEIWIT}.
 Besides the theoretical interest, there are also some interesting open
 questions concerning phenomenology.
 Although the most probable range of supersymmetry breaking mass
 parameters, as for instance the gluino mass, are at or above the 100
 GeV range, the possibility of a light gluino with a mass near 1 GeV
 is not yet completely excluded \cite{LIGHTGL}.
 In such a situation the extra hadronic states (``R-hadrons'') could be
 hidden in the complicated ``zoo'' of normal hadronic states with masses
 above 1 GeV \cite{FARRAR}.

 In a lattice formulation, most probably, one has to give up exact
 supersymmetry.
 This is certainly not surprising, because supersymmetry is an extension
 of the Poincar\'e symmetry, which is broken by the lattice.
 For instance, the supersymmetry generators 
 $Q_{i \alpha}, \bar{Q}_{j \dot{\beta}}$ with internal indices $i,j$ and
 Weyl-spinor indices $\alpha,\dot{\beta}$ ($i,j=1,..,N$; 
 $\alpha,\dot{\beta}=1,2$) obey the anticommutation relation
\be \label{eq01}
\left\{ Q_{i \alpha}, \bar{Q}_{j \dot{\beta}} \right\}
= 2\delta_{ij} \sigma_{\mu,\alpha\dot{\beta}} P_\mu \ , 
\ee
 where the four-momentum $P_\mu$ ($\mu=1,2,3,4$) is the generator of
 translation symmetry, which is broken on the latticized space-time.
 This led Curci and Veneziano \cite{CURVEN} to propose a lattice
 formulation of the $N=1$ supersymmetric Yang-Mills theory based on
 Wilson lattice fermions.
 The supersymmetry at zero gluino mass is then expected to be restored
 in the continuum limit, similarly to the axial symmetry in QCD at zero
 quark masses.
 The continuum limit at non-zero gluino mass is the corresponding theory
 with softly broken supersymmetry, which is actually the interesting
 case for phenomenology.

 The similar lattice formulation of $N=2$ supersymmetry \cite{N=2} is,
 on one hand, even simpler because there the gluino is a Dirac particle,
 in contrast to the $N=1$ case with Majorana gluinos.
 On the other hand, the $N=2$ supersymmetric Yang-Mills theory has
 some new generic features of supersymmetry, for instance, flat
 directions in the effective potential and asymptotically free Higgs
 mechanism.
 Theoretically very appealing is the automatically vector-like nature
 of N=2 SUSY, which is very welcome on the lattice, where the
 formulation of chiral gauge theories encounters great difficulties.

\section{AN ALGORITHM FOR GLUINOS}

 Although the $N=2$ extended supersymmetry has many interesting
 theoretical features, it is better to start the non-perturbative
 lattice studies with the $N=1$ case, which is simpler.

\subsection{Lattice action}

 The lattice action proposed in ref.~\cite{CURVEN} for the $N=1$
 supersymmetric Yang-Mills theory is based on the Wilson action
 formulated for a Dirac fermion in the adjoint representation.
 The fermionic part of this action is:
$$
S_f = \sum_x \LCB \overline{\psi}_x^r\psi_x^r 
-K \sum_{\mu=1}^4 \LSB
\overline{\psi}_{x+\hat{\mu}}^r V_{rs,x\mu}(1+\gamma_\mu)\psi_x^s
$$
\be \label{eq02}
+ \overline{\psi}_x^r V_{rs,x\mu}^T (1-\gamma_\mu)
\psi_{x+\hat{\mu}}^s \RSB \RCB \equiv \overline{\psi} Q \psi \ ,
\ee
 where the adjoint gauge field on links is
\be \label{eq03}
V_{rs,x\mu} \equiv V_{rs,x\mu}[U] \equiv
2 {\rm Tr}(U_{x\mu}^\dagger T_r U_{x\mu} T_s) \ ,
\ee
 and $T_r\; (r=1,2,3)$ denotes the SU(2) generators.
 The two Majorana field components of the Dirac fermion field $\psi$
 are defined as
\begin{eqnarray} 
\Psi^{(1)} & \equiv & \frac{1}{\sqrt{2}} ( \psi + C\overline{\psi}^T)\ ,
\nonumber \\ \label{eq04}
\Psi^{(2)} & \equiv & \frac{i}{\sqrt{2}} (-\psi + C\overline{\psi}^T)\ .
\end{eqnarray}
 In terms of these we have:
\be \label{eq05}
S_f = \half \overline{\Psi}^{(1)} Q \Psi^{(1)} 
    + \half \overline{\Psi}^{(2)} Q \Psi^{(2)} \ .
\ee
 Therefore a Majorana fermion corresponds to $N_f=\half$ number of
 flavours.
 As a consequence, in the bosonic path integral obtained by integrating
 out the Grassmann variables, the effect of a Majorana fermion is
 taken into account by the square root of the fermion determinant
 $\sqrt{\det Q}$.

\subsection{Two-step multi-bosonic algorithm}
 
 The numerical simulation of the $N=1$ supersymmetric Yang-Mills theory
 with dynamical gluinos requires the generation of $\sqrt{\det Q}$ in
 the bosonic path integral.
 Neglecting possible sign changes of $\det Q$, we can set with
 $\tilde{Q} \equiv \gamma_5 Q$
\be \label{eq06}
|\sqrt{\det(Q)}| = \{\det(Q^\dagger Q)\}^{1/4} 
= \{\det(\tilde{Q}^2)\}^{1/4} \ .
\ee
 In the framework of multi-bosonic algorithms \cite{LUSCHER}, these can
 be approximated by the determinant of polynomials.
 In order to decrease the number of pseudofermionic boson fields, and
 hence the autocorrelation and the storage requirements, one can take
 a two-step approximation scheme \cite{GLUINO}.
 The approximation up to a deviation norm $\delta$ is then
\be \label{eq07}
\{ \det \tilde{Q}^2 \}^{1/4} \;\stackrel{\delta}{\longrightarrow}\;
\frac{1}{\det[\bar{P}(\tilde{Q}^2)] \det[P(\tilde{Q}^2)]} \ .
\ee
 Here the polynomial $\bar{P}(x)$ is a crude approximation of the
 function $x^{-1/4}$ in the interval $x \in [\epsilon,\lambda]$
 containing the spectrum of $\tilde{Q}^2$.
 This approximation is then arbitrarily improved by the product of two
 polynomials $\bar{P}(x) P(x)$.
 In the fermion simulation algorithm \cite{GLUINO} the second factor
 $P(x)$ is taken into account by a ``noisy correction step'', as
 introduced in ref.~\cite{KENKUT}.
 (A similar correction scheme has been proposed in the multi-bosonic
 framework in ref.~\cite{BOFOGA}.)
 For the second polynomial one does not introduce bosonic fields, 
 therefore no storage is needed.
 The autocorrelation is mainly determined by the order of the first
 polynomial, which can be kept low.

 The definition of the deviation norm $\delta$ for the necessary 
 polynomial approximations has to allow for a flexibility in choosing
 functional forms, weight factors and approximation regions.
 This can be achieved by
\be \label{eq08}
\delta \equiv \left\{
\int_\epsilon^\lambda dx \left[ 1 - x^\alpha P(x) \right]^2
\right\}^\half \ ,
\ee
 where the function to be approximated is $x^{-\alpha}$.
 The advantage of this choice is that $\delta^2$ is a quadratic form
 of the polynomial coefficients and hence the minimum can easily be
 found.

 Compared to Chebyshev approximations, the polynomials minimizing
 $\delta$ in (\ref{eq08}) give somewhat less uniform relative deviations
 in the interval $[\epsilon,\lambda]$.
 (Remember that the Chebyshev polynomials are minimizing the maximum of
 the relative deviation, the so called ``infinity norm''.)
 However, in the present case the greater flexibility of the above
 quadratic scheme is more important.

\section{NUMERICAL SIMULATION}

 There are several interesting questions which can be investigated by
 numerical simulations in the $N=1$ supersymmetric Yang-Mills theory.
 For a first orientation the knowledge of the gluon-gluino-ball
 spectrum of hadronic states is very useful.
 In the supersymmetric limit there have to be degenerate
 supermultiplets.
 For non-zero gluino mass these multiplets are split up, and for
 infinite gluino mass only the glueballs of pure gauge theory remain in
 the physical spectrum.
 It is plausible (but not absolutely necessary) that the lowest $0^+$
 glueball state becomes a member of the lowest chiral supermultiplet,
 which also contains a Majorana spinor and a $0^-$ pseudoscalar boson.
 This latter can be most simply built from two gluinos.
 The Majorana spinor gluino-glueballs are obtained as bound states of
 a gluino with gluons.

\subsection{First results on the spectrum}

 The two-step multi-bosonic algorithm has several parameters which have
 to be optimized in a given situation (depending on bare parameters,
 lattices etc.).
 In ref.~\cite{GLUINO} a first look on autocorrelations has been taken
 for the gauge group SU(2) at bare gauge coupling $\beta=2.0$.
 (Another possible simulation algorithm based on classical dynamics
 type equations has recently been tested in ref.~\cite{DONGUA}.)
 For an interesting spectrum calculation one has to go, however, to
 larger $\beta$-values.

 In pure gauge theory the continuum behaviour starts to set in at
 $\beta=2.3$.
 Therefore one can try first this value.
 For hopping parameter values $K=0.14-0.15$ short autocorrelations
 could be achieved on $4^3 \cdot 8$ -- $8^3 \cdot 16$ lattices with
 the first polynomial $\bar{P}$ of order $\bar{n}=10-12$ and second
 polynomial $P$ of order $n=30-40$.
 The spectrum of $Q^\dagger Q$ was well inside the interval 
 $[\epsilon,\lambda]=[0.02,5.0]$.
\vspace*{-1.5em}
\begin{table}[h]
\caption{ \label{tab1}
 String tension $\sigma_2$, glueball mass $(M_{gg})$ and
 gluino-glueball mass $(M_{g\tilde{g}})$ in lattice units.}
\begin{center}
\begin{tabular}{|c|c|c|c|c|}
\hline
$\beta$  &  $K$  &
$\sqrt{\sigma_2}$  &  $M_{gg}$  &  $M_{g\tilde{g}}$  \\
\hline\hline
2.30  &  0.140  & 0.53(1) &  1.34(19)  &  2.90(12)  \\
\hline
2.30  &  0.150  & 0.54(1) &  0.85(23)  &  2.78(13)  \\
\hline\hline
\end{tabular}
\end{center}
\end{table}
\vspace*{-2.5em}

 The first results have been obtained on $4^3 \cdot 8$ lattice.
 The present statistics is based on 5000 updating cycles, where a
 cycle consisted of a few standard sweeps on every field variable.
 To monitor the string tension $\sigma$ I determined the $2 \otimes 2$
 Creutz-ratio $\chi_2 \equiv W_{22}W_{11}/W_{12}^2$ and defined
 $\sigma_2 \equiv -\log\chi_2 \simeq \sigma$.
 The results for this and the lowest $0^+$ gluball mass $(M_{gg})$ and
 gluino-glueball mass $(M_{g\tilde{g}})$ are contained in table
 \ref{tab1}.
 The signal for the gluino-gluino states can be obtained from noisy
 estimators \cite{GLUINO}, but this still needs some improvement, in
 order to obtain $M_{\tilde{g}\tilde{g}}$
 First estimates based on the study of the spectrum of $Q$ indicate
 a critical hopping parameter $K_c \simeq 0.165$.

\subsection{Future goals}

 Besides the spectrum, some other non-per\-turbative questions can be
 investigated which have also been recently discussed in the literature.
 Examples are: the phase structure based on exact $\beta$-functions
 \cite{KOGSHI}, or the question of confinement versus screening at zero
 gluino mass \cite{GKMS}.
 These problems seem to be well suited for numerical simulation studies
 which are feasible with presently available computer resources.


\end{document}